\documentclass{SciPost}

\hypersetup{
    colorlinks,
    linkcolor={red!50!black},
    citecolor={blue!50!black},
    urlcolor={blue!80!black}
}

\usepackage[bitstream-charter]{mathdesign}
\usepackage{dsfont}
\DeclareSymbolFont{usualmathcal}{OMS}{cmsy}{m}{n}
\DeclareSymbolFontAlphabet{\mathcal}{usualmathcal}

\fancypagestyle{SPstyle}{
\fancyhf{}
\lhead{\colorbox{scipostblue}{\bf \color{white} ~SciPost Physics }}
\rhead{{\bf \color{scipostdeepblue} ~Submission }}

\fancyfoot[C]{\textbf{\thepage}}
}

\begin{document}

\pagestyle{SPstyle}

\begin{center}{\Large \textbf{\color{scipostdeepblue}{
Soliton-antisoliton pairs in the supersymmetric gapped phase of an interacting Majorana chain\\
}}}\end{center}

\begin{center}\textbf{
Alberto Nocera\textsuperscript{1},
Mobin Shakeri\textsuperscript{1},
Armin Rahmani\textsuperscript{2$\star$} and
Ian Affleck\textsuperscript{1}
}\end{center}

\begin{center}
{\bf 1} Department of Physics and Astronomy and Stewart Blusson Quantum Matter Institute,University of British Columbia, Vancouver, B.C., Canada, V6T 1Z1
\\
{\bf 2} Department of Physics and Astronomy and Advanced Materials Science and Engineering Center, Western Washington University, Bellingham, Washington 98225, USA
\\[\baselineskip]
$\star$ \href{mailto:armin.rahmani@wwu.edu}{\small armin.rahmani@wwu.edu}
\end{center}

\section*{\color{scipostdeepblue}{Abstract}}
\textbf{\boldmath{
A strongly interacting chain of Majorana fermions realizes the supersymmetric tricritical Ising phase, with supersymmetry (SUSY) extending into a symmetry-broken ordered phase adjacent to the tricritical point. Although the signatures of SUSY at the tricritical point are well understood, their behavior in the gapped phase remains less clear. Here, we address two key questions: how SUSY manifests in the gapped phase and what is the nature of the excitations in this phase. We show that, in the thermodynamic limit, a conventional SUSY diagnostic that remains finite at the tricritical point diverges immediately on the Ising side, yet decays continuously to zero deeper in the gapped phase, signaling the persistence of SUSY. Focusing on the lowest excited states in the supersymmetric gapped regime, we find that the excitations consist of soliton–antisoliton pairs separating distinct ordered regions. Each soliton binds an emergent localized Majorana mode, and together the pair forms a nonlocal Dirac fermion. The occupation of this Dirac mode distinguishes eigenstates with even and odd fermion parity.
}}

\vspace{\baselineskip}

\noindent\textcolor{white!90!black}{%
\fbox{\parbox{0.975\linewidth}{%
\textcolor{white!40!black}{\begin{tabular}{lr}%
  \begin{minipage}{0.6\textwidth}%
    {\small Copyright attribution to authors. \newline
    This work is a submission to SciPost Physics. \newline
    License information to appear upon publication. \newline
    Publication information to appear upon publication.}
  \end{minipage} & \begin{minipage}{0.4\textwidth}
    {\small Received Date \newline Accepted Date \newline Published Date}%
  \end{minipage}
\end{tabular}}
}}
}


\vspace{10pt}
\noindent\rule{\textwidth}{1pt}
\tableofcontents
\noindent\rule{\textwidth}{1pt}
\vspace{10pt}


\begin{center}
\begin{minipage}{0.9\textwidth}
\itshape
{\textit {\small This paper is dedicated to Ian Affleck, a unique physicist and exceptional mentor. Ian was instrumental in generating the ideas that led to this work, and we were fortunate to collaborate with him in the beginning of the project. He remained very interested in seeing this work completed. His insights, generosity, and unwavering support have not only influenced this research, but have also profoundly shaped our careers and scientific development.}}
\end{minipage}
\end{center}

\section{Introduction}
Lattice models exhibiting supersymmetry (SUSY) have been studied for many years~\cite{Fendley2003,Lee2007,Huijse2011,Jian2015}. More recently, interacting Majorana fermions have emerged as a promising route to realizing spacetime SUSY in condensed matter systems~\cite{Grover2014,Ponte2014}. Strongly interacting Majorana zero modes may be experimentally accessible in magnetic-field-induced vortex lattices on the surface of topological superconductors, including the Fu--Kane platform of a proximitized topological insulator~\cite{Fu2008,Biswas2013,Mishmash2019,Chiu2015}. These developments have stimulated extensive interest in interacting Majorana lattice models~\cite{Affleck2017, Rahmani2019a, Rahmani2019b,Li2018, Tummuru2021}. Several one-dimensional lattice models involving only Majorana building blocks have been shown to realize the supersymmetric tricritical Ising (TCI) model at the interface between a critical Ising phase and a symmetry-broken gapped phase~\cite{Rahmani2015,Obrien2018}. The degeneracy of the gapped phase depends on boundary conditions (BC), but with periodic BC on the Majorana fermions, corresponding to the Ramond sector of the field theory, the phase is two-fold degenerate~\cite{Rahmani2015, Rahmani2015a}. While supersymmetry is spontaneously broken in the critical Ising phase, it survives in the adjacent gapped phase~\cite{Friedan1985,Zamolodchikov1986}.

In this paper, we investigate numerical signatures of supersymmetry (SUSY) in the gapped phase and explore the nature of its low-energy excitations. Focusing on the first excited states, we present evidence that the excitation energy arises from a soliton–antisoliton (SA) pair that separates regions corresponding to the two symmetry-breaking orders of the degenerate ground states. The SA pair, however, is not localized at fixed positions, and the excited states are therefore superpositions of configurations with the pair at different locations. We find that each soliton and antisoliton binds a localized Majorana mode. With periodic BC on the Majorana chain, the spectrum is at least doubly degenerate, since each state has a partner with opposite fermion parity. The key distinction between excited states with even and odd total fermion parity lies in the occupation of the Dirac fermion formed by these two localized Majorana modes at the soliton and antisoliton locations.

The remainder of this paper is organized as follows. In Sec.~\ref{sec:susy}, we introduce the model and numerically examine the fate of supersymmetry as we move away from the supersymmetric TCI point. In Sec.~\ref{sec:dimer}, we introduce a dimerization order parameter for the two symmetry-broken ground states of the gapped phase. Section~\ref{sec:sa} focuses on the detection of SA pairs in the first excited states. In Sec.~\ref{sec:maj}, we explore the emergent Majorana modes bound to solitons and antisolitons. Finally, conclusions are presented in Sec.~\ref{sec:conc}. Details on a level crossing in response to uniform and staggered order-parameter fields are provided in the appendix.

\section{Model, supercharges, and SUSY away from the TCI point}
\label{sec:susy}

\begin{figure}[t]
    \includegraphics[width=.63\linewidth]{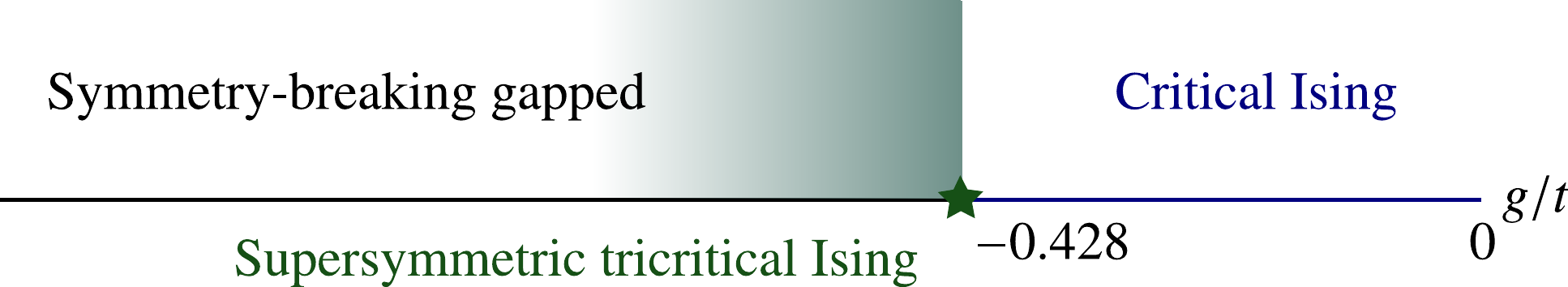}
    \centering
    \caption{The phase diagram of the OF model of Eq.~\eqref{eq:hamil} as a function of $g/t$ for $g<0$.}
    \label{fig:1}
\end{figure}

In this paper, we focus on the O'Brien–Fendley (OF) model~\cite{Obrien2018} of interacting Majorana fermions, which realizes the TCI point at an interaction strength of order one, unlike the Majorana-Hubbard chain with interactions among four consecutive sites~\cite{Rahmani2015}, which realizes the TCI point at an interaction strength of around 250 (See Ref.~\cite{Aasen2020} for an explanation). The supersymmetric TCI state of the OF model was recently created in the IBM quantum computer for finite system sizes~\cite{Samanta}. The Hamiltonian of the model is
\begin{equation}\label{eq:hamil}
H_0=\sum_{j=0}^{N-1} it\gamma_j\gamma_{j+1}+g\gamma_{j-2}\gamma_{j-1}\gamma_{j+1}\gamma_{j+2},
\end{equation}
where $t$ and $g$ respectively represent the hybridization amplitude and interaction strength between Majoranas. The fermion parity $P=\prod_{j=0}^{N/2-1} (i\gamma_{2j+1}\gamma_{2j})$ is conserved by this Hamiltonian. We primarily focus on the Ramond sector of the model with periodic boundary conditions $\gamma_{j+N}\sim \gamma_j$ on the Majorana fermions, which corresponds to either periodic or antiperiodic boundary conditions on the spins when the model is mapped to spin-$1/2$ variables via a Jordan--Wigner transformation, depending on the fermion parity. The phase diagram of this model for negative $g$ is shown in Fig.~\ref{fig:1}. The TCI point separates a critical Ising phase with central charge $c=1/2$~\cite{Friedan1985,Cappelli1987} from a gapped phase with broken symmetry. The tricritical point is described by a conformal field theory with central charge $c=7/10$, which is known to exhibit superconformal symmetry~\cite{Rahmani2015a}.

At the field-theory level, SUSY is described by two commuting fermionic supercharge operators $Q_{\rm FT}$ and $\bar Q_{\rm FT}$, with the Hamiltonian given by $H_{\rm FT}=(Q_{\rm FT})^2+(\bar Q_{\rm FT})^2$. These operators convert fermionic excitations to bosonic excitations, and vice versa. The densities of these supercharges are known as supercurrents $T_{\rm FT}$ and $\bar T_{\rm FT}$, which are primary fields with conformal dimensions $(3/2,0)$ and $(0,3/2)$, respectively.

The lattice versions of the supercharge and supercurrents have been identified for Hamiltonian~\eqref{eq:hamil} in Refs.~\cite{Obrien2018,Zou2020}. Although the lattice versions of the supercharges do not commute, these lattice operators flow to the field-theoretic operators under renormalization. The lattice supercurrent operators are given by \cite{Obrien2018,Zou2020}
\begin{align}\label{eq:supercurrent}
T_{j}&\propto t(\gamma_{2j-1}+\gamma_{2j})-2ig(\gamma_{2j-2}+\gamma_{2j+1})\gamma_{2j-1}\gamma_{2j},\\
\bar{T}_{j}&\propto t(\gamma_{2j-1}-\gamma_{2j})-2ig (\gamma_{2j+1}-\gamma_{2j-2})\gamma_{2j-1}\gamma_{2j}.
\end{align}
with lattice supercharge operators given by $Q=\sum_j T_j$ and $\bar Q=\sum_j \bar T_j$.

Utilizing the mapping between fermionic and bosonic states generated by the supercharges, we can use appropriate matrix elements of $Q$ as a diagnostic for SUSY, as identified by Zhou and Vidal~\cite{Zou2020}. We note that the operator content in the Ramond sector (periodic boundary conditions on Majoranas) contains four primary fields $(\sigma, \sigma', \mu, \mu')$, with respective chiral conformal dimensions $h$ given by $(3/80, 7/16, 3/80, 7/16)$. These fields have zero conformal spin, so the chiral dimensions $h$ and $\bar h$ are equal. The superconformal algebra then implies that the states corresponding to these primary operators transform as follows:
\begin{align}
Q|\sigma\rangle&={\cal C} \sqrt{h_\sigma-c/24}|\mu\rangle={{\cal C} \over \sqrt{120}}|\mu\rangle,\\
Q|\sigma'\rangle&={\cal C} \sqrt{h_{\sigma'}-c/24}|\mu'\rangle={7{\cal C} \over \sqrt{120}}|\mu'\rangle.
\end{align}

For a lattice model with periodic boundary conditions, the states $|\sigma\rangle$ and $|\mu\rangle$ are respectively the ground states in the sector with even and odd fermion parity, while $|\sigma'\rangle$ and $|\mu'\rangle$ are the first excited states in the same fermion parity sectors. We then define the ratio $R$, which is equal to $1/7$ at the TCI point, and use its behavior to track the fate of SUSY as we move away from the TCI point:
\begin{equation}\label{eq:r}
    R=|{\langle \mu|Q|\sigma\rangle} /{\langle \mu'|Q|\sigma'\rangle }|.
\end{equation}

The salient property of $Q$ is that it switches the fermion parity while preserving the ordering of states in the spectrum. Thus, a finite value of $R$ in the thermodynamic limit can serve as a diagnostic for the survival of SUSY away from the superconformal point. A value of $R$ equal to zero or infinity indicates that applying $Q$ produces a state of opposite parity that is orthogonal to its partner in the other parity sector, respectively for the ground and first-excited states.

\begin{figure}[t]
    \includegraphics[width=.9\linewidth]{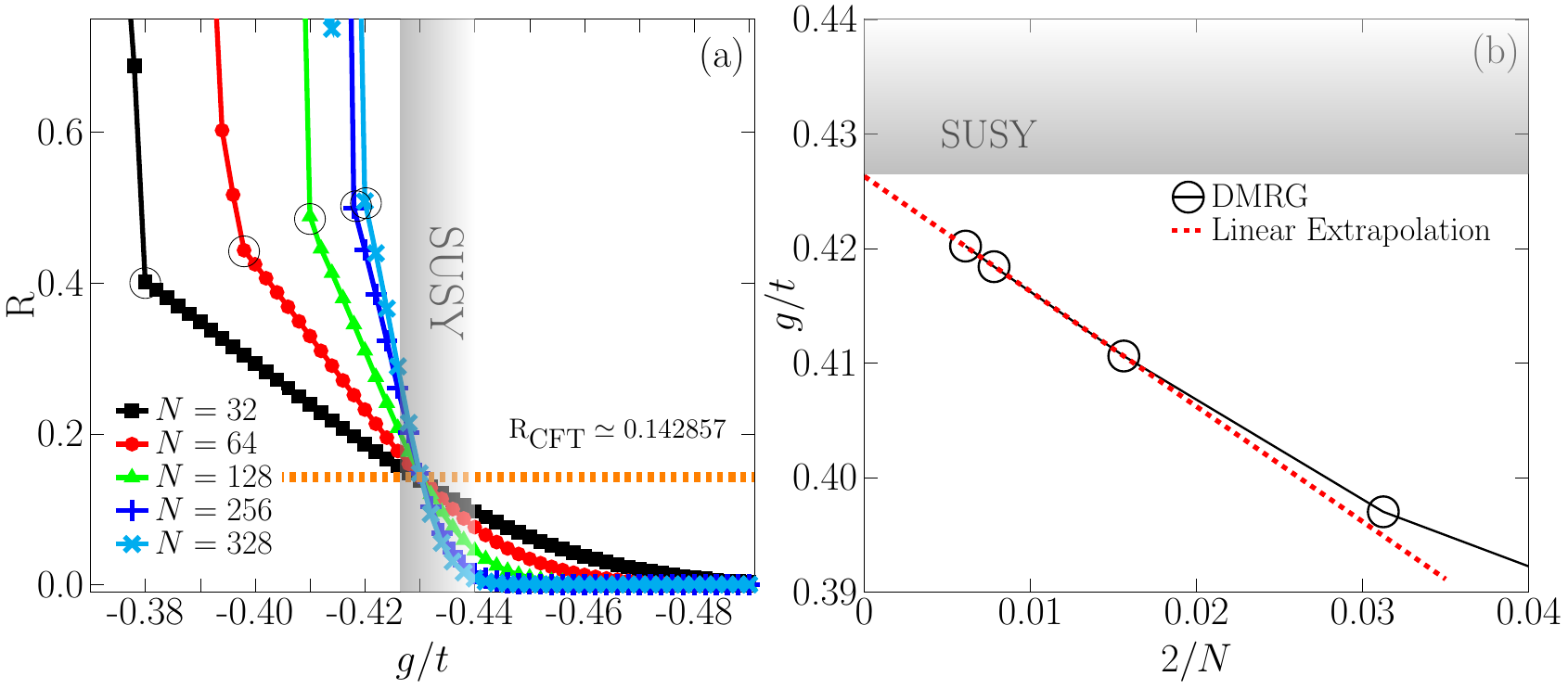}
    \centering
    \caption{\label{fig:2}(a) The numerically computed value of $R$ defined in Eq.~\eqref{eq:r} as a function of $g/t$ for various system sizes. A kink appears on the Ising side, which extrapolates toward the TCI point. (b) Finite-size scaling of the position of the kink.}
\end{figure}

In Fig.~\ref{fig:2}(a), we show the numerically calculated (with DMRG) values of $R$ for multiple system sizes as a function of $g/t$. The TCI point was predicted to occur at $g/t\approx -0.428$ in Ref.~\cite{Obrien2018} using various energy gap ratios. It appears that for $g/t=-0.428$, $R$ extrapolates to a larger value than $1/7$, and the value of $g/t$ that gives rise to $R=1/7$ in the thermodynamic limit lies slightly to the left of $-0.428$. This apparent discrepancy was already observed in Ref.~\cite{Zou2020}, which suggested that the parameters $t$ and $g$ in the supercurrent~\eqref{eq:supercurrent} might need to be considered as variational parameters instead of the bare Hamiltonian parameters. Despite the small uncertainty in the precise position of the TCI point, these results provide numerical evidence for the spontaneous breaking of SUSY on the Ising side of the transition, and the survival of SUSY in the gapped phase in the vicinity of the TCI point.

As shown in Fig.~\ref{fig:2}(a), to the right of the transition, $R$ exhibits a kink and a jump to much larger values as we move away from the TCI point and into the Ising phase. This dramatic change in the behavior of $R$, i.e., the position of this kink, shifts toward the TCI point upon increasing the system size. In Fig.~\ref{fig:2}(b), we show the finite-size scaling of the position of the kink, which indicates its extrapolation to the TCI point in the thermodynamic limit. Thus, in the thermodynamic limit, $R$ appears to diverge when moving to the Ising phase on the right-hand side of the TCI point, signaling spontaneous breaking of SUSY. However, on the gapped side, $R$ changes with a finite derivative at the TCI point, indicating the survival of SUSY in the vicinity of the TCI point and its gradual disappearance deep in the gapped phase, where $R$ decays to zero.

The calculation of $R$ away from the TCI point and the pronounced difference in its extrapolated behavior at large system sizes provide a numerical diagnostic for the breaking of SUSY on the Ising side and its survival in the gapped phase in the vicinity of the TCI point.
\section{Dimerization order and associated symmetry-breaking fields}\label{sec:dimer}

\begin{figure}[t]
\vspace{4mm}
\begin{center}
\includegraphics[width=.45\linewidth]{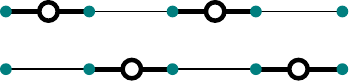}
\end{center}
\caption{Schematic representation of the ordering of the two degenerate states as the vacuum for two different sets of Dirac fermions (thick bonds) constructed from neighboring Majoranas (green circles).}
\label{fig:3}
\end{figure}

In this section, we discuss the nature of the ordering in the gapped phase. At the effective field-theory level, the OF model captures the same physics as the Rahmani-Zhu-Franz-Affleck (RZFA) model, which has the Majorana interaction term on four consecutive sites instead of four sites $j \pm 1$ and $j \pm 2$ in a cluster of five consecutive sites. Based on the strong-coupling limit $g \to \infty$, the two ordered states in the RZFA model for $t > 0$ correspond to empty Dirac sites for two patterns of pairing neighboring Majoranas into Dirac fermions, as shown in Fig.~\ref{fig:3}. Because the two models are known to display the same low-energy behavior, exhibiting a TCI transition from a gapless $c = 1/2$ Ising phase to a doubly degenerate ordered gapped phase, we expect the nature of the ground-state ordering in the gapped phase to be similar for both models.

We now specify an order parameter, motivated by the strong-coupling regime, which we use throughout the ordered gapped phase. The order parameter vanishes at the TCI transition. The ideal states shown in Fig.~\ref{fig:3} can be considered the vacua $|0\rangle_c$ and $|0\rangle_d$ for Dirac fermions:
\begin{equation}
c_j = \gamma_{2j} + i\gamma_{2j+1}, \quad d_j = \gamma_{2j-1} + i\gamma_{2j}.
\end{equation}

We can write all Majorana operators in terms of the $c$ Dirac fermions as
\begin{equation}
\gamma_{2j} = (c_j + c^\dagger_j)/2, \qquad \gamma_{2j+1} = (c_j - c^\dagger_j)/(2i).
\end{equation}

The relationship above implies that the $d$ fermions are not independent and can be written as
\begin{equation}\label{eq:dc}
d_j = \frac{i}{2}\left(c_j + c^\dagger_j - c_{j-1} + c^\dagger_{j-1}\right).
\end{equation}

We note that the fermion parity defined below Eq.~\eqref{eq:hamil} can be written in terms of the occupation numbers of these Dirac fermions as
\begin{equation}\label{eq:parity}
P = \prod_{j=1}^L (1 - 2n_c^j) = -\prod_{j=1}^L (1 - 2n_d^j),
\end{equation}
where $L = N/2$ is the number of Dirac fermions.

The expectation value of $n_c^j = c^\dagger_j c_j$ with respect to the vacuum $|0\rangle_c$ is identically zero, i.e., $\langle n_c^j\rangle_c = 0$. Similarly, we have $\langle n_d^j\rangle_d = 0$, where $n_d^j$ is the occupation number operator for the $d$ fermions. Using Eq.~\eqref{eq:dc}, we find $\langle n_d^j\rangle_c = 1/2$. Moreover, translation symmetry implies $\langle n_c^j\rangle_d = 1/2$. Thus, the difference between the expectation values of $n_c$ and $n_d$, namely $\langle n_c\rangle - \langle n_d\rangle$, is equal to $\pm 1/2$ for the two dimerized states, and its absolute value can serve as the order parameter for this ordered phase.

The two ordered states shown in Fig.~\ref{fig:3} occur in different parity sectors, with the top (bottom) configuration in Fig.~\ref{fig:3} corresponding to the even (odd) parity sector. This is due to our definition in Eq.~\eqref{eq:parity}, noting that the top (bottom) panel is the vacuum of $c$ ($d$) fermions. Since the two states are degenerate, an infinitesimal dimerization field coupled to $n_c - n_d$ can lift the degeneracy between the two orderings, and consequently between the odd and even parity sectors for $\Delta = 0$, selecting one of the individual dimerized states as the unique ground state. The corresponding Hamiltonian perturbation for small $\Delta$ can be written as
\begin{equation}\label{eq:uniform}
H' = \Delta \sum_j (n^j_c - n^j_d)
= \sum_j 2i\Delta(\gamma_{2j}\gamma_{2j+1} - \gamma_{2j-1}\gamma_{2j}),
\end{equation}
with $H = H_0 + H'$.

\begin{figure}[t]
\vspace{4mm}
\centering
\includegraphics[width=.81\linewidth]{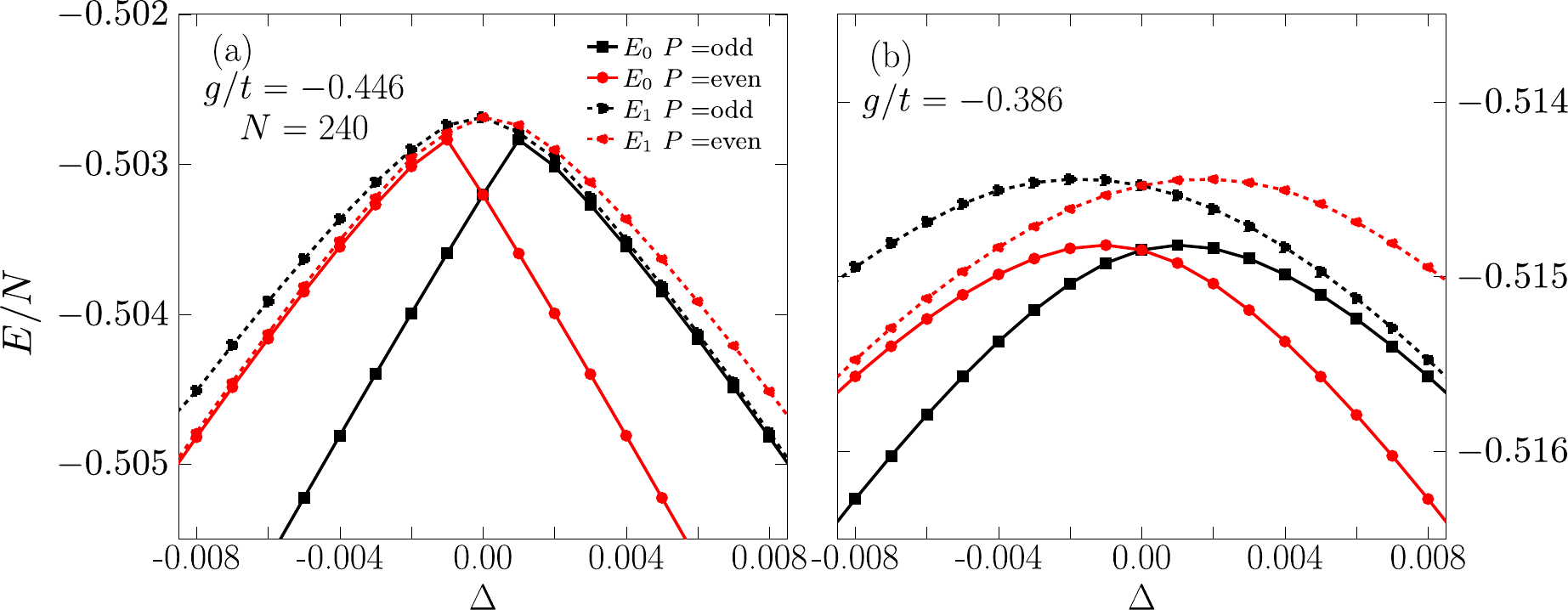}
\caption{\label{fig:4} The energy densities of the ground and first-excited states in the two fermion parity sectors for $g/t$ in the ordered phase [panel (a)] and in the Ising phase [panel (b)]. The ground-state energy exhibits a linear dependence on $\Delta$ in the ordered phase for small $\Delta$. A positive (negative) $\Delta$ lowers the energy in the even (odd) fermion parity sector, leading to a unique ground state with even (odd) parity. These results were obtained with the energy scale $t$ set to one.}
\end{figure}

In Fig.~\ref{fig:4}(a), the energies of the ground state and the first excited state are plotted in the two fermion parity sectors. In the ordered phase, for small $\Delta$, the ground state exhibits a linear dependence on $\Delta$ (unlike the first excited state), confirming the nature of the order parameter. The slope of the ground-state energy as a function of $\Delta$ is positive (negative) for odd (even) fermion parity. For larger $\Delta$, there is a level crossing where the nature of the ground state changes from the two dimerized states discussed above, and the linear dependence is lost. In the Ising phase, as shown in Fig.~\ref{fig:4}(b), we do not observe a linear dependence on $\Delta$ or a level crossing.

We have observed this behavior across multiple system sizes and expect it to persist in the thermodynamic limit. However, the order in which the limits of small $\Delta$ and large $N$ are taken is subtle due to the finite-size scaling of the level-crossing position discussed above. Since we are interested in the system’s response to an infinitesimal $\Delta$, it is necessary to take the limit $\Delta \to 0$ \textit{before} the thermodynamic limit. This order of limits is crucial because, as discussed in Appendix~\ref{app:scaling}, the value $\Delta_{\rm LC}$, where the level-crossing occurs, decreases with increasing system size. Given the limitations of numerical finite-size scaling, it remains difficult to determine whether $\Delta_{\rm LC}$ vanishes or saturates to a small finite value as $N \to \infty$. Nevertheless, the simple illustrative argument presented in Appendix~\ref{app:scaling} suggests that the scenario $\Delta_{\rm LC} \to 0$ is plausible and provides a consistent picture, even though it would imply a discontinuity in the energy in the thermodynamic limit, since the system remains gapped at $\Delta = 0$.

The spectrum shown in Fig.~\ref{fig:4} exhibits an important symmetry $E_{\rm even}(\Delta) = E_{\rm odd}(-\Delta)$ with periodic boundary conditions. This relation is a direct consequence of translation by one Majorana site. The Hamiltonian $H_0 + H'$ is invariant under the simultaneous transformations $\gamma_j \to \gamma_{j+1}$ and $\Delta \to -\Delta$. However, the transformation $\gamma_j \to \gamma_{j+1}$ changes the fermion parity $P$ due to the Majorana anticommutation relations~\cite{Rahmani2015a, Hsieh2016} [see Eq.~\eqref{eq:parity}, where $c$ and $d$ are exchanged under a single-site translation].

\begin{figure}[t]
\centering
\includegraphics[width=.81\linewidth]{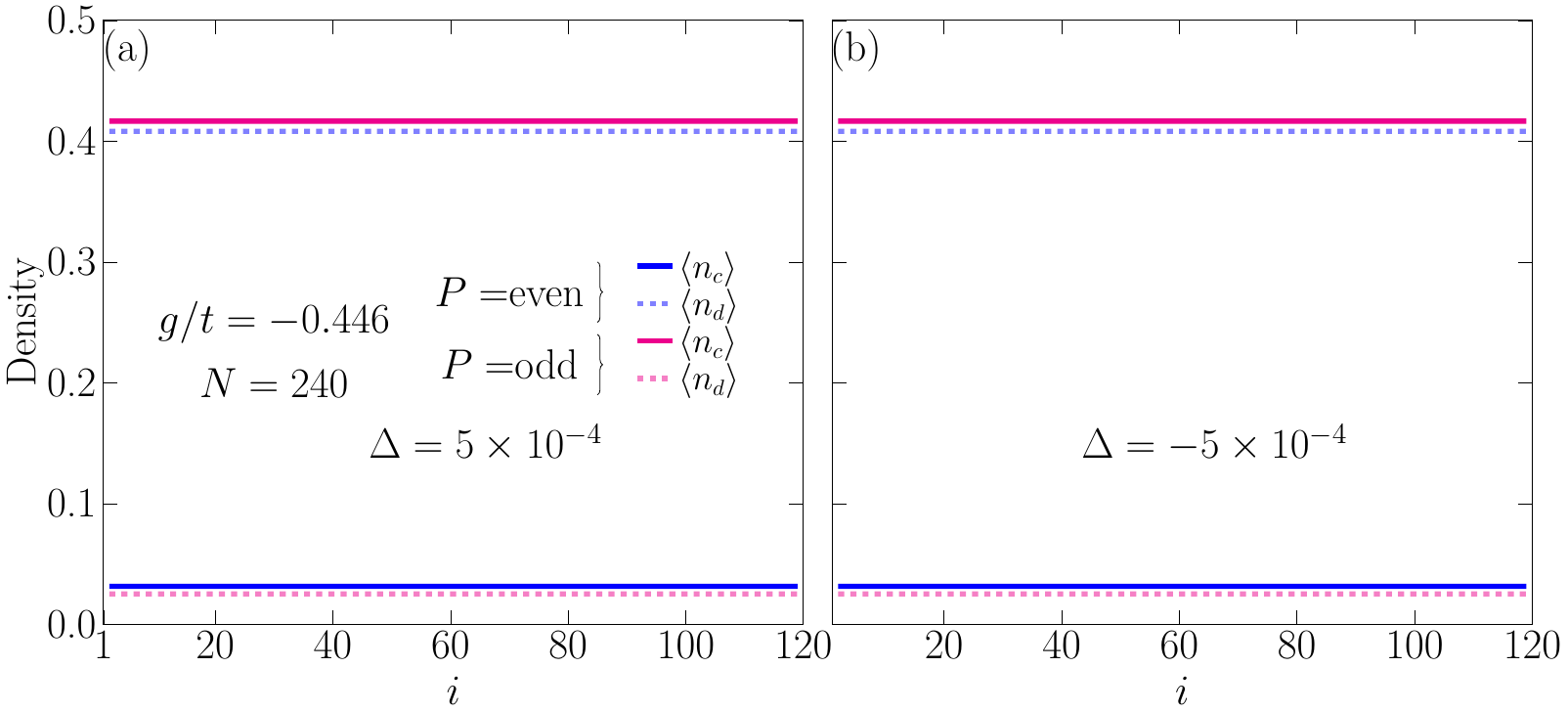}
\caption{\label{fig:5} Density of Dirac fermions for positive $\Delta$ (panel a) and negative $\Delta$ (panel b) in each parity sector. The value of $|\Delta|$ has been taken smaller than the level crossing shown in Fig.~\ref{fig:4}(a). Regardless of the sign of $\Delta$, the even and odd parity sectors have $\langle n_c\rangle \simeq 0$, $\langle n_d\rangle \simeq 1/2$ and $\langle n_c\rangle \simeq 1/2$, $\langle n_d\rangle \simeq 0$, respectively.}
\end{figure}

We now examine the ground-state expectation values of $n_c$ and $n_d$ in the presence of the symmetry-breaking perturbation $H'$. As shown in Fig.~\ref{fig:5}(a) for a system of $160$ Majoranas, the nature of the order is determined entirely by the parity sector. Although the uniform perturbation can raise or lower the energy, it does not change the nature of the state within a given parity sector. For both positive and negative $\Delta$ (as well as $\Delta = 0$), the even fermion-parity sector is characterized by $\langle n_c\rangle \simeq 0$ (much smaller than $1/2$) and $\langle n_d\rangle \simeq 1/2$. In the odd sector, the opposite ordering is favored, with $\langle n_d\rangle \simeq 0$ and $\langle n_c\rangle \simeq 1/2$.

\section{Soliton--antisoliton pair in the first excited states}
\label{sec:sa}

Since the ground state is doubly degenerate, corresponding to two distinct dimerization patterns, the lowest-energy excitations are naturally expected to consist of a small number of solitons and antisolitons separating domains of the two orders~\cite{Jackiw1976,Su1979}. In particular, the first excited state should be dominated by a single soliton--antisoliton (SA) pair, since additional pairs would incur a higher energy cost. There is, however, no reason to expect the positions of the soliton and antisoliton, or even their separation, to be fixed. Rather, the eigenstates are expected to be quantum superpositions of configurations with SA pairs located at different positions along the chain. Consequently, the expectation values $\langle n_{c,d}\rangle$ with respect to eigenstates of $H_0$ are not useful for detecting individual SA pairs. Indeed, translation symmetry forces these expectation values to be spatially uniform, thereby obscuring the localized domain-wall structure associated with the soliton and antisoliton.

\begin{figure}[t]
\vspace{4mm}
\centering
\includegraphics[width=.95\linewidth]{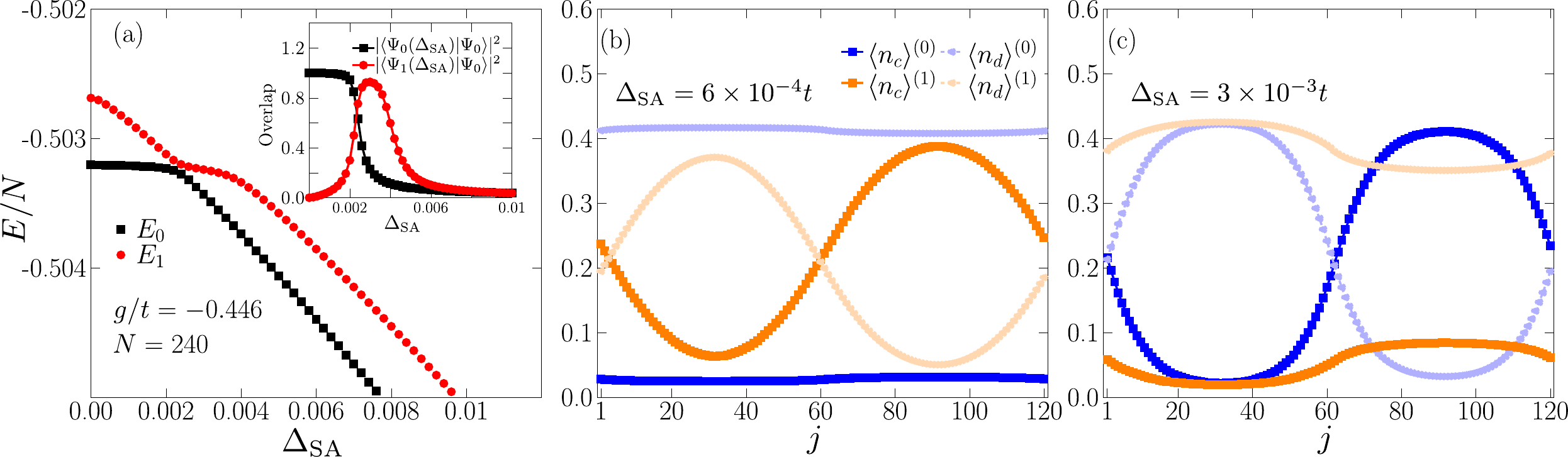}
\caption{\label{fig:6}
Panel (a): Ground and first excited state energies as a function of $\Delta_{\rm SA}$ (with SA distance $m = L/2$). Inset: wave-function overlaps as a function of $\Delta_{\rm SA}$, showing clear evidence of a level crossing. Panels (b--c): Density of Dirac fermions for values of $\Delta_{\rm SA}$ below and above the level crossing observed in panel (a).
}
\end{figure}

We can, however, construct a perturbation to which states containing SA pairs exhibit a strong susceptibility. The perturbation in Eq.~\eqref{eq:sa} explicitly favors one of the two dimerization patterns associated with the degenerate ground states. Therefore, applying this perturbation with $\Delta = \Delta_{\rm SA}$ in a segment of the chain with $a < m$, and with $\Delta = -\Delta_{\rm SA}$ in the complementary region $a \geqslant m$, energetically favors opposite dimerization patterns on the two sides of the interface. As a result, an SA pair is expected to become pinned near the boundaries between these regions, namely around $a \approx 0$ and $a \approx m$. We therefore write the perturbation corresponding to the SA pinning field $\Delta_{\rm SA}$ as
\begin{equation}\label{eq:sa}
H_{\rm SA} = \sum_{a} 2i\Delta_{\rm SA}s_a(\gamma_{2a}\gamma_{2a+1} - \gamma_{2a-1}\gamma_{2a}),
\end{equation}
where $s_a = \pm 1$ for $a < m$ and $a \geqslant m$.

One comment is in order regarding fermion parity. Although the uniform ordered ground states without SA pairs are distinguished and protected by fermion-parity symmetry, this does not preclude the local coexistence of the two ordering patterns within an excited state. Fermion parity is a global symmetry, and therefore domains with different local dimerization patterns can appear in excited states in both fermion parity sectors.

We then use the susceptibility to the SA-pinning perturbation introduced above as a diagnostic of the extent to which the unperturbed excited state is dominated by soliton--antisoliton pairs. Figure~\ref{fig:6}(a) shows the energies of the ground state and first excited state in the even-parity sector; the odd-parity sector displays qualitatively similar behavior. We have used $m = L/2$ so that the positive and negative $\Delta_{\rm SA}$ are each applied to one half of the chain.

As shown in Fig.~\ref{fig:6}(a), the ground-state energy remains nearly unchanged for small values of $\Delta_{\rm SA}$. This behavior is expected because the ground state contains a uniform dimerization pattern throughout the chain. The SA-pinning perturbation therefore raises the energy in one half of the system while lowering it by a comparable amount in the other half, resulting in only a weak net effect. In contrast, the energy of the first excited state exhibits a pronounced dependence on $\Delta_{\rm SA}$, providing direct evidence that this state contains a significant SA-pair component.

We note that, as in the case of the uniform perturbation, a level crossing occurs in any finite system. This crossing is inherited from the corresponding level crossing in the uniform case and therefore does not reflect a distinct physical mechanism. As before, the physically relevant order of limits is to take $\Delta_{\rm SA} \to 0$ prior to the thermodynamic limit. Consequently, for finite-size systems, we restrict our analysis to values of $\Delta_{\rm SA}$ that remain below the critical value at which the level crossing occurs. Beyond the level crossing between the ground and first excited states, the roles of the two states are reversed. The excited-state energy becomes nearly independent of $\Delta_{\rm SA}$, while the ground-state energy develops a strong dependence on the perturbation. This interpretation is corroborated by the overlaps of the perturbed ground and first excited states with the unperturbed ground state $|\Psi_0\rangle$, shown in the inset. At the crossing, the two states effectively exchange character, as evidenced by the abrupt increase in the overlap of $|\Psi_1(\Delta_{\rm SA})\rangle$ with $|\Psi_0\rangle$ and the corresponding decrease in the overlap of $|\Psi_0(\Delta_{\rm SA})\rangle$. For larger values of $\Delta_{\rm SA}$, the system exits the perturbative regime. Both overlaps decrease substantially, and the energies of both states acquire an approximately linear dependence on $\Delta_{\rm SA}$. In this regime, the perturbation no longer acts merely as a probe of preexisting SA pairs; instead, it qualitatively modifies the states by nonperturbatively inducing soliton--antisoliton pairs.

The pinning of the SA pair in the first excited state can be seen more directly from the expectation values of $n_c$ and $n_d$. As shown in Fig.~\ref{fig:6}(b) for a value of $\Delta_{\rm SA}$ below the level-crossing point, the ground-state expectation values undergo only minor changes between the left and right halves of the chain, consistent with the absence of an SA pair. In contrast, the first excited state displays a clear signature of a pinned SA pair, with the local ordering pattern changing across the interfaces and attaining the order favored by the perturbation in both halves of the chain.

This behavior persists beyond the level crossing, although the roles of the ground and first excited states are interchanged. As shown in Fig.~\ref{fig:6}(c), the state that acquires the strong response to the SA-pinning field continues to exhibit the characteristic signatures of a pinned SA pair, while the other state remains only weakly affected. This exchange of behavior reflects the interchange of the two eigenstates at the level crossing discussed above.

\begin{figure}[t]
\centering
\includegraphics[width=.9\linewidth]{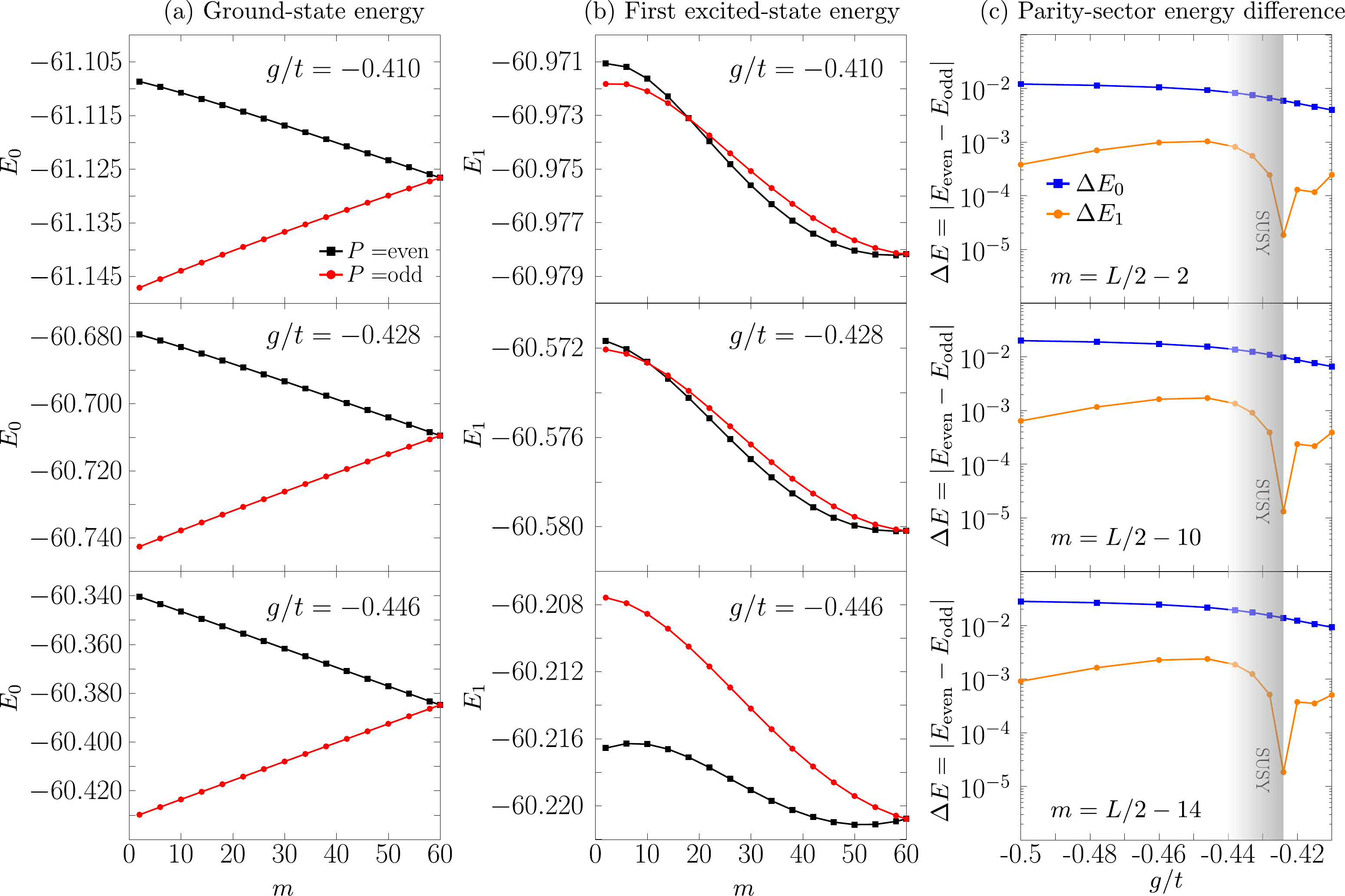}
\caption{\label{fig:7}
(a--b) Ground-state and first excited state energies in the two fermion parity sectors as a function of the soliton location $m$. All data are for $L = 120$ Dirac fermions and SA pinning field $\Delta_{\rm SA} = 0.001$. The top, middle, and bottom panels correspond respectively to the Ising phase ($g/t = -0.410$), the TCI point ($g/t = -0.428$), and the gapped phase ($g/t = -0.446$). (c) The difference $\Delta E_{0,1} = |E^{0,1}_{\rm even} - E^{0,1}_{\rm odd}|$ between ground states (first excited states) of the two parity sectors at fixed SA locations $m$ close to the maximum $m = L/2$ as a function of $g/t$.
}
\end{figure}

So far, we have applied the SA pinning field~\eqref{eq:sa} at $m = L/2$, probing soliton--antisoliton pairs separated by half the chain length. We now examine pinning fields at other separations. Figures~\ref{fig:7}(a,b) summarize the evolution of the low-energy spectrum in the presence of the $\Delta_{\rm SA}$ pinning field for three representative interaction strengths corresponding to the Ising regime near the TCI point (top), the TCI point itself (middle), and the gapped phase near the TCI point (bottom). Panel (a) shows the ground-state energies in the two parity sectors, while panel (b) shows the corresponding first excited-state energies, both as a function of the SA separation.

When the positive and negative pinning fields are applied to exactly half of the chain ($m = L/2$), the ground states in the two parity sectors remain degenerate. Deviating from $m = L/2$ introduces an imbalance between the positive and negative pinning fields, favoring one type of order over the other. As a result, the ground-state energies exhibit the same linear dependence that appeared in the uniform-field case, now as a function of $m$ at fixed $\Delta_{\rm SA}$. This is because the ground states for both parity sectors correspond to uniform order without any SA pairs. The behavior of the first excited states is more subtle. Although they are also degenerate at $m = L/2$, varying $m$ generates a splitting between the parity sectors. This splitting, however, is substantially smaller than that observed for the ground states.

Interestingly, the splitting between the first excited states in the two parity sectors, $\Delta E_1$, measured at fixed $m$ near $L/2$ as a function of $g/t$, provides a useful diagnostic of the TCI point. Although both $\Delta E_0$ and $\Delta E_1$ vanish when $m = L/2$, they become finite once the soliton--antisoliton separation is shifted away from this balanced configuration. While the ground-state splitting $\Delta E_0$ is relatively insensitive to $g/t$, the first-excited-state splitting exhibits a pronounced suppression, approaching near degeneracy at the TCI point. The behavior of $\Delta E_1$ therefore serves as a sensitive probe of the changing nature of the low-energy excitations across the TCI transition. These results are shown in Fig.~\ref{fig:7}(c), where the top, middle, and bottom panels correspond to $L/2 - m = 2$, $10$, and $14$, respectively.

\section{Emergent Majoranas at solitons and anti-solitons}
\label{sec:maj}

In the ground state, the two types of order correspond to the two fermion-parity sectors. The picture that emerges from the previous section suggests that the first excited state can be viewed as a superposition of states containing a single SA pair separating the two types of order. In this regime, the connection between fermion parity, which is a global property, and the order type appears to break down, since both types of order are energetically favorable.

\begin{figure}[t]
\begin{center}
    \includegraphics[width=.36\linewidth]{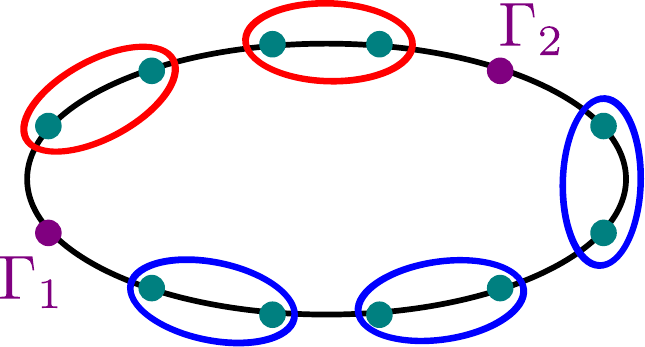}
\end{center}
\centering
\caption{A cartoon illustration of a direct-product state containing an SA pair (see Ref.~\cite{Aasen2020} for a more detailed discussion). Switching between the two types of order, represented by the red and blue ovals and corresponding to different pairings of Majorana modes (green circles) into Dirac fermions, leaves two Majorana modes, $\Gamma_1$ and $\Gamma_2$ (purple circles), decoupled. The fermion parity of the full state is then determined solely by the parity operator $i\Gamma_1\Gamma_2$ associated with the Dirac fermion formed from these two modes bound to the soliton and antisoliton.}
\label{fig:8}
\end{figure}

As predicted in Ref. ~\cite{Aasen2020}, for a product state with a single SA pair, the cartoon picture of Fig.~\ref{fig:8} suggests the emergence of two Majorana modes, each localized at a domain wall, in excited states that contain a single SA pair. A domain wall between the two ordered patterns shown in Fig.~\ref{fig:8} corresponds to a change in the pairing of Majorana modes into Dirac fermions. As the pairing switches between the red and blue patterns, two Majorana modes, $\Gamma_1$ and $\Gamma_2$, remain unpaired and localize near the soliton and antisoliton. These modes form a nonlocal Dirac fermion, so that the fermion parity of the full state is determined solely by the operator $i\Gamma_1\Gamma_2$. 

In the general case, the emergent Majorana modes are not identical to the microscopic Majorana operators on the lattice, but instead can be written as linear combinations of them:
\begin{equation}
\Gamma_1 = \sum_n \alpha_n \gamma_n,
\qquad
\Gamma_2 = \sum_n \beta_n \gamma_n.
\end{equation}
The coefficients satisfy $\sum_n \alpha_n^2 = \sum_n \beta_n^2 = 1$ and $\sum_n \alpha_n \beta_n = 0$, in order to preserve the correct anticommutation relations. If these emergent Majorana modes are bound to the soliton and antisoliton, we expect the coefficients $\alpha_n$ and $\beta_n$ to be localized near the corresponding domain walls, decaying exponentially or faster (e.g., approximately Gaussian).

If the even- and odd-fermion-parity sectors are determined solely by the occupation of the nonlocal Dirac mode formed from $\Gamma_1$ and $\Gamma_2$, then we must have $i\langle \Gamma_1 \Gamma_2 \rangle = \pm 1$
in the two fermion-parity sectors. Due to the local nature of the emergent Majorana modes in terms of the original lattice Majoranas, we expect their presence to manifest in the Majorana Green's functions
\begin{equation}
G^{\mathrm{even,odd}}_{n,m} =
i\langle \gamma_n \gamma_m \rangle_{\mathrm{even,odd}}.
\end{equation}
In particular, the difference between the even- and odd-parity sectors should exhibit localized signatures near the soliton and antisoliton. A localized difference between Majorana two-point functions for even and odd fermion-parity sectors near the solitons and antisolitons would suggest that there may be two Majorana modes bound to these defects whose fermion parity determines the parity of the entire state.

We note that, in the OF model, the Green's function vanishes whenever both Majorana operators belong to the same sublattice, namely
\begin{equation}\label{eq:sublattice}
    G^{\mathrm{even,odd}}_{2j, 2k} = G^{\mathrm{even,odd}}_{2j+1, 2k+1} = 0, \quad k \neq j.
\end{equation}
This is a consequence of spatial reflection symmetry. Let us relabel the sites such that the midpoint between two sites belonging to the same sublattice is denoted as site $0$. Under the reflection transformation, $\gamma_j \to (-1)^j \gamma_{-j}$, the Hamiltonian remains invariant for periodic BC, while the Majorana Green's function changes sign. Consequently, symmetry requires $G_{nm}=0$ whenever $n$ and $m$ belong to the same sublattice, as in Eq.~\eqref{eq:sublattice}.

\begin{figure}[t]
\begin{center}
    \includegraphics[width=.95\linewidth]{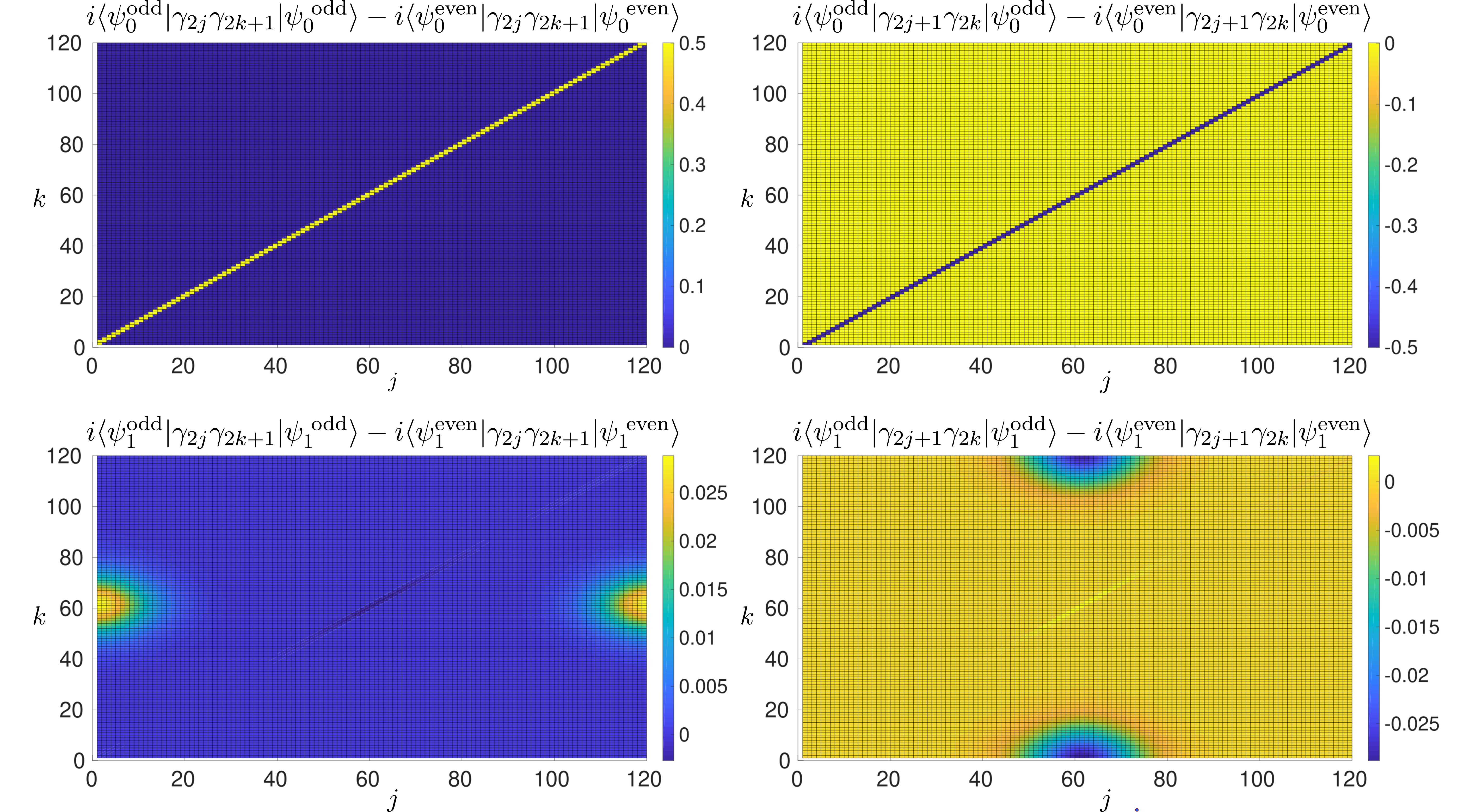}
\end{center}
\centering
\caption{\label{fig:9}Difference of Majorana Green’s functions between even and odd fermion-parity sectors. Top panels show the ground state, where the difference vanishes except for diagonal components $j=k$, consistent with the ordered phases corresponding to the $c$ and $d$ Dirac vacua. Bottom panels show the first excited state with a pinned soliton–antisoliton pair, where the difference is dominated by contributions localized near the SA pair. In this figure, $L=120$ Dirac fermions ($N=240$) and SA pinning field $\Delta_{SA}=0.0006t$.}
\end{figure}

We now visualize the Majorana Green’s function for the soliton–antisoliton configuration. A weak pinning field $\Delta_{\rm SA}$, applied below the level crossing, is used to stabilize an SA pair at separation $m = L/2$. We compute all Majorana Green’s functions and verify that those connecting Majoranas on the same sublattice (even–even and odd–odd) vanish. Accordingly, we focus on the even–odd and odd–even correlators, which are plotted separately in Fig.~\ref{fig:9}.

It is particularly instructive to examine the difference between these correlators in the even and odd fermion-parity sectors. For the ground state, this difference vanishes everywhere except for the diagonal components $j=k$, consistent with the expected structure of the two ordered phases, which serve as the vacua of the $c$ and $d$ Dirac fermions, as shown in the top panels of Fig.~\ref{fig:9} for both even–odd and odd–even Majorana sectors. The bottom panels display the same quantities for the first excited states in the presence of pinned SA pairs. The difference between the two fermion-parity sectors is primarily concentrated near the SA-pair locations, consistent with the emergence of Majorana modes localized at these positions.

\section{Conclusions}
\label{sec:conc}
In this work, we investigated the persistence of SUSY and the nature of low-energy excitations in the gapped phase adjacent to the supersymmetric tricritical Ising point of the OF model of an interacting Majorana chain. While the realization of emergent SUSY at the tricritical point is well established, much less has been understood about how supersymmetric structures manifest away from criticality. Our results provide numerical evidence that SUSY survives over a finite region of the ordered phase.

Using matrix elements of the lattice supercharge, we examined the behavior of the ratio $R$ introduced in Ref.~\cite{Zou2020}. Finite-size scaling indicates a striking asymmetry on the two sides of the TCI point. On the Ising side, $R$ appears to diverge upon crossing the transition, consistent with spontaneous SUSY breaking. In contrast, on the ordered side, $R$ evolves smoothly away from its tricritical value and decreases continuously toward zero only deep in the gapped phase. This behavior provides a useful numerical diagnostic for distinguishing the supersymmetric and nonsupersymmetric regions of the phase diagram.

We further characterized the ordered phase through a dimerization order parameter constructed from two inequivalent pairings of neighboring Majorana fermions into Dirac modes. An infinitesimal field coupled to this order parameter selects one of the two degenerate ordered states as the unique ground state and produces a linear response of the ground-state energy. Interestingly, in the absence of the field, the two degenerate ordered states are distinguished by their fermion parity and are therefore protected from mixing by this global symmetry. The role of the symmetry-breaking field is not to alter the nature of these states, but rather to lift their degeneracy by lowering the energy of one fermion-parity sector while raising the energy of the other.

A central result of this work is the identification of the lowest excited states as states dominated by a single SA pair separating regions with the two distinct dimerization patterns. Because translationally invariant eigenstates are superpositions of domain-wall configurations, local observables do not directly spatially identify the solitons. We therefore introduced an SA-pinning field that locally favors opposite orders in different halves of the chain. The strong susceptibility of the first excited state to this perturbation, together with the resulting spatial profile of the order parameter, provides direct evidence that the low-energy excitation contains an SA pair. The response of the ground state, by contrast, remains weak, consistent with its uniform ordering.

The soliton picture naturally leads to the emergence of localized Majorana modes bound to the domain walls. In the strong-coupling limit, the change in dimerization pattern leaves an unpaired Majorana mode at each defect. Our numerical results support the continuation of this picture throughout the supersymmetric gapped regime. The two bound Majorana modes form a nonlocal Dirac fermion whose occupation distinguishes the even- and odd-parity sectors. Consequently, the parity quantum number of the excited state can be understood as arising from an emergent fermionic degree of freedom associated with the SA pair rather than from the bulk ordering itself.

These results reveal a rich structure in the supersymmetric gapped phase of the OF model. The emergence of soliton-bound Majorana modes and their connection to fermion parity provide a microscopic picture of the excitation spectrum and suggest new ways in which supersymmetric physics can manifest away from criticality. More broadly, our results demonstrate that important signatures of emergent SUSY survive beyond the critical point and continue to organize the low-energy physics throughout a finite region of the ordered phase.

Several interesting questions remain open. While our results provide a microscopic picture of the lowest excitations in the supersymmetric gapped phase, it would be valuable to extend this understanding to higher-energy states and to develop a more direct description of the SA excitations and their bound Majorana modes. It would also be interesting to determine whether SUSY imposes additional constraints on the dynamics of these domain walls. More broadly, our work raises the possibility that analogous soliton-bound fermionic modes and persistent signatures of emergent SUSY may occur in other lattice realizations of SUSY, including higher-dimensional systems.

\section*{Acknowledgments}
We thank Will Holdhusen, Sutapa Smanta, and Jian-Xin Zhu for helpful discussions. We also thank Marcel Franz and Dmitry Pikulin for valuable collaborations on related projects. A.R. acknowledges support from the U.S. Department of Energy under Grant No. DE-SC0024641.

\begin{appendix}
\numberwithin{equation}{section}

\section{Order of limits for the level crossings}
\label{app:scaling}
We mentioned in Secs.~\ref{sec:dimer} and \ref{sec:sa} that the position of the level crossing appears to approach a zero field (or at least a very small field) if we take the thermodynamic limit first. Therefore, we must first take the limit of small fields used to pin the soliton and antisoliton in a finite system and then take the thermodynamic limit.

This behavior is rather strange as it implies a discontinuity in energy as a function of $\Delta$ or $\Delta_{\rm SA}$, in the thermodynamic limit. While in any finite system, perturbation theory implies that the energy must change continuously, we can indeed have jumps in energy for an infinite system. A cartoon picture based on classical Ising chain demonstrates this behavior.
\begin{figure}[]
\begin{center}
    \includegraphics[width=.63\linewidth]{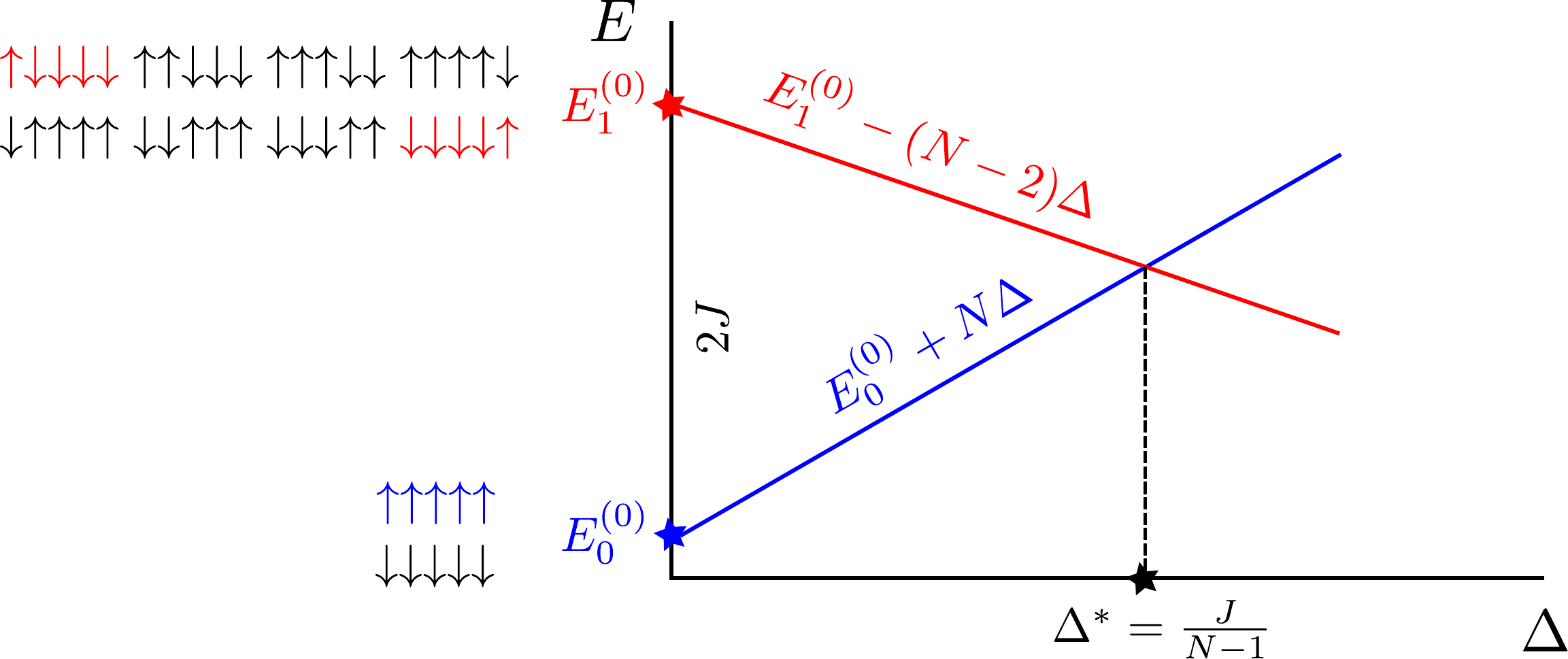}
    \end{center}
	\centering
	\caption{\label{fig:10}Energy levels of a ferromagnetic Ising chain with $N=5$ spins and open boundary conditions in the presence of a uniform symmetry-breaking field $H'=\sum_i \Delta \sigma_i^z$. One of the two degenerate ordered ground states, with all spins up, whose energy increases with $\Delta$, is shown in blue. Two of the eight first-excited states containing a single up spin, whose energies decrease most rapidly with $\Delta$, are shown in red. The field induces a level crossing between the blue and red states at $\Delta^*=J/(N-1)$, which extrapolates to zero in the thermodynamic limit even though the states are separated by a gap of $2J$ for $\Delta=0$ in any finite system.}
\end{figure}

Consider a classical ferromagnetic Ising chain of $N$ spins with Hamiltonian 
\begin{equation}
  H=-\sum_{i=1}^{N-1}J\sigma_i^z\sigma_{i+1}^z.  
\end{equation}
We use open boundary conditions as an example. The ground state is two-fold degenerate with energy $E_0^{(0)}=-(N-1)J$, all spins up or down. The first excited states have degeneracy $2(N-1)$ as we can place a domain wall on any of the $N-1$ bonds, leading to an energy $E_0^{(1)}=-(N-3)J$, with a gap of $2J$. Now consider a perturbation comprised of a field coupled to the order parameter $H'=\sum_{i=1}^N\Delta \sigma_i^z$. This perturbation increases the energy of the all-up state, and maximally decreases the energy of the two excited states with $N-1$ down spins. In Fig. \ref{fig:10}, this unpertubed ground state and the excited states above are respectively shown in blue and red for $N=5$. The energies of these states then become $E_0^{(0)}+N\Delta$ and $E_1^{(0)}-(N-2)\Delta$, leading to a level crossing at $\Delta^*=J/(N-1)$. We then observe that the critical field corresponding to the level crossing extrapolates to zero in the thermodynamic limit. For any finite system, however, the field pick the all-down state as the ground state, making the all-down state the first excited state and the twofold-degenerate states with $N-1$ down spins as the second excited states, with ground and first excited states becoming degenerate as $\Delta\to 0$, remaining separated from the higher-energy levels by a gap $2J$.
We note that if we apply a field analogous to $\Delta_{\rm SA}$ of the form  
\begin{equation}\label{eq:delta}
    H'=\sum_{i=1}^{N/2}\Delta_{\rm SA}\sigma_i^z-\sum_{i=N/2+1}^{N}\Delta_{\rm SA}\sigma_i^z,
\end{equation}
to this chain for even $N$, the twofold-degenerate all-up and all-down states remain degenerate with the same unperturbed energy $E_0^{(0)}$. Among the degenerate first excited states, however, a single configuration with $N/2$ down (up) spins on the left (right) half of the chain experiences the maximal reduction in energy, acquiring an energy $E_0^{(0)}-N\Delta_{\rm SA}$. This leads to a level crossing at $\Delta_{\rm SA}^*=\frac{2J}{N}$, which is inherited from the level crossing in the uniform-field case and similarly shifts toward $\Delta_{\rm SA}=0$ in the thermodynamic limit. The behavior shown in Fig.~\ref{fig:6}(a) likely reflects a similar mechanism.

\end{appendix}






\bibliography{susy.bib}

\begin{thebibliography}{10}
\providecommand{\url}[1]{\texttt{#1}}
\providecommand{\urlprefix}{URL }
\expandafter\ifx\csname urlstyle\endcsname\relax
  \providecommand{\doi}[1]{doi:\discretionary{}{}{}#1}\else
  \providecommand{\doi}{doi:\discretionary{}{}{}\begingroup
  \urlstyle{rm}\Url}\fi
\providecommand{\eprint}[2][]{\url{#2}}

\bibitem{Fendley2003}
P.~Fendley, K.~Schoutens and J.~de~Boer,
\newblock \emph{Lattice models with n=2 supersymmetry},
\newblock Phys. Rev. Lett. \textbf{90}, 120402 (2003),
\newblock \doi{10.1103/PhysRevLett.90.120402}.

\bibitem{Lee2007}
S.-S. Lee,
\newblock \emph{Emergence of supersymmetry at a critical point of a lattice
  model},
\newblock Phys. Rev. B \textbf{76}, 075103 (2007),
\newblock \doi{10.1103/PhysRevB.76.075103}.

\bibitem{Huijse2011}
L.~Huijse, N.~Moran, J.~Vala and K.~Schoutens,
\newblock \emph{Exact ground states of a staggered supersymmetric model for
  lattice fermions},
\newblock Phys. Rev. B \textbf{84}, 115124 (2011),
\newblock \doi{10.1103/PhysRevB.84.115124},
\newblock \eprint{1103.1368}.

\bibitem{Jian2015}
S.-K. Jian, Y.-F. Jiang and H.~Yao,
\newblock \emph{Emergent spacetime supersymmetry in 3d weyl semimetals and 2d
  dirac semimetals},
\newblock Phys. Rev. Lett. \textbf{114}, 237001 (2015),
\newblock \doi{10.1103/PhysRevLett.114.237001}.

\bibitem{Grover2014}
T.~Grover, D.~N. Sheng and A.~Vishwanath,
\newblock \emph{Emergent space-time supersymmetry at the boundary of a
  topological phase},
\newblock Science \textbf{344}, 280 (2014),
\newblock \doi{10.1126/science.1248253}.

\bibitem{Ponte2014}
P.~Ponte and S.-S. Lee,
\newblock \emph{Emergence of supersymmetry on the surface of three-dimensional
  topological insulators},
\newblock New Journal of Physics \textbf{16}(1), 013044 (2014),
\newblock \doi{10.1088/1367-2630/16/1/013044}.

\bibitem{Fu2008}
L.~Fu and C.~L. Kane,
\newblock \emph{Superconducting proximity effect and majorana fermions at the
  surface of a topological insulator},
\newblock Phys. Rev. Lett. \textbf{100}, 096407 (2008),
\newblock \doi{10.1103/PhysRevLett.100.096407}.

\bibitem{Biswas2013}
R.~R. Biswas,
\newblock \emph{Majorana fermions in vortex lattices},
\newblock Phys. Rev. Lett. \textbf{111}, 136401 (2013),
\newblock \doi{10.1103/PhysRevLett.111.136401}.

\bibitem{Mishmash2019}
R.~V. Mishmash, A.~Yazdani and M.~P. Zaletel,
\newblock \emph{Majorana lattices from the quantized hall limit of a
  proximitized spin-orbit coupled electron gas},
\newblock Phys. Rev. B \textbf{99}, 115427 (2019),
\newblock \doi{10.1103/PhysRevB.99.115427}.

\bibitem{Chiu2015}
C.-K. Chiu, D.~I. Pikulin and M.~Franz,
\newblock \emph{Strongly interacting majorana fermions},
\newblock Phys. Rev. B \textbf{91}, 165402 (2015),
\newblock \doi{10.1103/PhysRevB.91.165402}.

\bibitem{Affleck2017}
I.~Affleck, A.~Rahmani and D.~Pikulin,
\newblock \emph{Majorana-hubbard model on the square lattice},
\newblock Phys. Rev. B \textbf{96}, 125121 (2017),
\newblock \doi{10.1103/PhysRevB.96.125121}.

\bibitem{Rahmani2019a}
A.~Rahmani, D.~Pikulin and I.~Affleck,
\newblock \emph{Phase diagrams of majorana-hubbard ladders},
\newblock Phys. Rev. B \textbf{99}, 085110 (2019),
\newblock \doi{10.1103/PhysRevB.99.085110}.

\bibitem{Rahmani2019b}
A.~{Rahmani} and M.~{Franz},
\newblock \emph{{Interacting Majorana fermions}},
\newblock Reports on Progress in Physics \textbf{82}(8), 084501 (2019),
\newblock \doi{10.1088/1361-6633/ab28ef},
\newblock \eprint{1811.02593}.

\bibitem{Li2018}
C.~Li and M.~Franz,
\newblock \emph{Majorana-hubbard model on the honeycomb lattice},
\newblock Phys. Rev. B \textbf{98}, 115123 (2018),
\newblock \doi{10.1103/PhysRevB.98.115123}.

\bibitem{Tummuru2021}
T.~Tummuru, A.~Nocera and I.~Affleck,
\newblock \emph{Triangular lattice majorana-hubbard model: Mean-field theory
  and dmrg on a width-4 torus},
\newblock Phys. Rev. B \textbf{103}, 115128 (2021),
\newblock \doi{10.1103/PhysRevB.103.115128}.

\bibitem{Rahmani2015}
A.~Rahmani, X.~Zhu, M.~Franz and I.~Affleck,
\newblock \emph{Emergent supersymmetry from strongly interacting majorana zero
  modes},
\newblock Phys. Rev. Lett. \textbf{115}, 166401 (2015),
\newblock \doi{10.1103/PhysRevLett.115.166401}.

\bibitem{Obrien2018}
E.~O'Brien and P.~Fendley,
\newblock \emph{Lattice supersymmetry and order-disorder coexistence in the
  tricritical ising model},
\newblock Phys. Rev. Lett. \textbf{120}, 206403 (2018),
\newblock \doi{10.1103/PhysRevLett.120.206403}.

\bibitem{Rahmani2015a}
A.~Rahmani, X.~Zhu, M.~Franz and I.~Affleck,
\newblock \emph{Phase diagram of the interacting majorana chain model},
\newblock Phys. Rev. B \textbf{92}, 235123 (2015),
\newblock \doi{10.1103/PhysRevB.92.235123}.

\bibitem{Friedan1985}
D.~Friedan, Z.~Qiu and S.~Shenker,
\newblock \emph{Superconformal invariance in two dimensions and the tricritical
  ising model},
\newblock Physics Letters B \textbf{151}(1), 37 (1985),
\newblock \doi{https://doi.org/10.1016/0370-2693(85)90819-6}.

\bibitem{Zamolodchikov1986}
A.~B. Zamolodchikov,
\newblock \emph{Conformal symmetry and multicritical points in two-dimensional
  quantum field theory},
\newblock Sov. J. Nucl. Phys. \textbf{44}, 529 (1986).

\bibitem{Aasen2020}
D.~Aasen, R.~S.~K. Mong, B.~M. Hunt, D.~Mandrus and J.~Alicea,
\newblock \emph{Electrical probes of the non-abelian spin liquid in kitaev
  materials},
\newblock Phys. Rev. X \textbf{10}, 031014 (2020),
\newblock \doi{10.1103/PhysRevX.10.031014}.

\bibitem{Samanta}
S.~Samanta, J.-X. Zhu and A.~Rahmani,
\newblock \emph{Realizing supersymmetry in a digitized quantum device} (2025),
  \eprint{2504.18703}.

\bibitem{Cappelli1987}
A.~Cappelli,
\newblock \emph{Modular invariant partition functions of superconformal
  theories},
\newblock Physics Letters B \textbf{185}(1), 82 (1987),
\newblock \doi{https://doi.org/10.1016/0370-2693(87)91532-2}.

\bibitem{Zou2020}
Y.~Zou and G.~Vidal,
\newblock \emph{Emergence of conformal symmetry in quantum spin chains:
  Antiperiodic boundary conditions and supersymmetry},
\newblock Phys. Rev. B \textbf{101}, 045132 (2020),
\newblock \doi{10.1103/PhysRevB.101.045132}.

\bibitem{Hsieh2016}
T.~H. Hsieh, G.~B. Hal\'asz and T.~Grover,
\newblock \emph{All majorana models with translation symmetry are
  supersymmetric},
\newblock Phys. Rev. Lett. \textbf{117}, 166802 (2016),
\newblock \doi{10.1103/PhysRevLett.117.166802}.

\bibitem{Jackiw1976}
R.~Jackiw and C.~Rebbi,
\newblock \emph{Solitons with fermion number 1/2},
\newblock Phys. Rev. D \textbf{13}, 3398 (1976),
\newblock \doi{10.1103/PhysRevD.13.3398}.

\bibitem{Su1979}
W.~P. Su, J.~R. Schrieffer and A.~J. Heeger,
\newblock \emph{Solitons in polyacetylene},
\newblock Phys. Rev. Lett. \textbf{42}, 1698 (1979),
\newblock \doi{10.1103/PhysRevLett.42.1698}.

\end{thebibliography}


\end{document}